\documentstyle[twoside,epsfig]{article}

\catcode`\@=11
\long\def\@makefntext#1{
\protect\noindent \hbox to 3.2pt {\hskip-.9pt  
$^{{\eightrm\@thefnmark}}$\hfil}#1\hfill}		

\def\@makefnmark{\hbox to 0pt{$^{\@thefnmark}$\hss}}	
	
\def\ps@myheadings{\let\@mkboth\@gobbletwo
\def\@oddhead{\hbox{}
\rightmark\hfil\eightrm\thepage}   
\def\@oddfoot{}\def\@evenhead{\eightrm\thepage\hfil
\leftmark\hbox{}}\def\@evenfoot{}
\def\sectionmark##1{}\def\subsectionmark##1{}}

\newcommand{\LL}{\ell^+ \ell^-}

\oddsidemargin=\evensidemargin
\newcounter{sectionc}\newcounter{subsectionc}\newcounter{subsubsectionc}
\renewcommand{\section}[1] {\vspace{12pt}\addtocounter{sectionc}{1} 
\setcounter{subsectionc}{0}\setcounter{subsubsectionc}{0}\noindent 
	{\tenbf\thesectionc. #1}\par\vspace{5pt}}
\renewcommand{\subsection}[1] {\vspace{12pt}\addtocounter{subsectionc}{1} 
	\setcounter{subsubsectionc}{0}\noindent 
	{\bf\thesectionc.\thesubsectionc. {\kern1pt \bfit #1}}\par\vspace{5pt}}
\renewcommand{\subsubsection}[1] {\vspace{12pt}\addtocounter{subsubsectionc}{1}
	\noindent{\tenrm\thesectionc.\thesubsectionc.\thesubsubsectionc.
	{\kern1pt \tenit #1}}\par\vspace{5pt}}
\newcommand{\nonumsection}[1] {\vspace{12pt}\noindent{\tenbf #1}
	\par\vspace{5pt}}

\newcounter{appendixc}
\newcounter{subappendixc}[appendixc]
\newcounter{subsubappendixc}[subappendixc]
\renewcommand{\thesubappendixc}{\Alph{appendixc}.\arabic{subappendixc}}
\renewcommand{\thesubsubappendixc}
	{\Alph{appendixc}.\arabic{subappendixc}.\arabic{subsubappendixc}}

\renewcommand{\appendix}[1] {\vspace{12pt}
        \refstepcounter{appendixc}
        \setcounter{figure}{0}
        \setcounter{table}{0}
        \setcounter{lemma}{0}
        \setcounter{theorem}{0}
        \setcounter{corollary}{0}
        \setcounter{definition}{0}
        \setcounter{equation}{0}
        \renewcommand{\thefigure}{\Alph{appendixc}.\arabic{figure}}
        \renewcommand{\thetable}{\Alph{appendixc}.\arabic{table}}
        \renewcommand{\theappendixc}{\Alph{appendixc}}
        \renewcommand{\thelemma}{\Alph{appendixc}.\arabic{lemma}}
        \renewcommand{\thetheorem}{\Alph{appendixc}.\arabic{theorem}}
        \renewcommand{\thedefinition}{\Alph{appendixc}.\arabic{definition}}
        \renewcommand{\thecorollary}{\Alph{appendixc}.\arabic{corollary}}
        \renewcommand{\theequation}{\Alph{appendixc}.\arabic{equation}}
        \noindent{\tenbf Appendix \theappendixc #1}\par\vspace{5pt}}
\newcommand{\subappendix}[1] {\vspace{12pt}
        \refstepcounter{subappendixc}
        \noindent{\bf Appendix \thesubappendixc. {\kern1pt \bfit #1}}
	\par\vspace{5pt}}
\newcommand{\subsubappendix}[1] {\vspace{12pt}
        \refstepcounter{subsubappendixc}
        \noindent{\rm Appendix \thesubsubappendixc. {\kern1pt \tenit #1}}
	\par\vspace{5pt}}

\newcommand{\textlineskip}{\baselineskip=13pt}
\newcommand{\smalllineskip}{\baselineskip=10pt}

\def\eightcirc{
\begin{picture}(0,0)
\put(4.4,1.8){\circle{6.5}}
\end{picture}}
\def\eightcopyright{\eightcirc\kern2.7pt\hbox{\eightrm c}} 

\newcommand{\copyrightheading}[1]
	{\vspace*{-2.5cm}\smalllineskip{\flushleft
	{\footnotesize International Journal of Modern Physics A, #1}\\
	{\footnotesize $\eightcopyright$\, World Scientific Publishing
	 Company}\\
	 }}

\def\abstracts#1#2#3{{
	\centering{\begin{minipage}{4.5in}\baselineskip=10pt\footnotesize
	\parindent=0pt #1\par 
	\parindent=15pt #2\par
	\parindent=15pt #3
	\end{minipage}}\par}} 

\renewenvironment{thebibliography}[1]
	{\frenchspacing
	 \ninerm\baselineskip=11pt
	 \begin{list}{\arabic{enumi}.}
	{\usecounter{enumi}\setlength{\parsep}{0pt}
	 \setlength{\leftmargin 12.7pt}{\rightmargin 0pt} 
	 \setlength{\itemsep}{0pt} \settowidth
	{\labelwidth}{#1.}\sloppy}}{\end{list}}

\newcounter{itemlistc}
\newcounter{romanlistc}
\newcounter{alphlistc}
\newcounter{arabiclistc}

\newcommand{\fcaption}[1]{
        \refstepcounter{figure}
        \setbox\@tempboxa = \hbox{\footnotesize Fig.~\thefigure. #1}
        \ifdim \wd\@tempboxa > 5in
           {\begin{center}
        \parbox{5in}{\footnotesize\smalllineskip Fig.~\thefigure. #1}
            \end{center}}
        \else
             {\begin{center}
             {\footnotesize Fig.~\thefigure. #1}
              \end{center}}
        \fi}

\newcommand{\tcaption}[1]{
        \refstepcounter{table}
        \setbox\@tempboxa = \hbox{\footnotesize Table~\thetable. #1}
        \ifdim \wd\@tempboxa > 5in
           {\begin{center}
        \parbox{5in}{\footnotesize\smalllineskip Table~\thetable. #1}
            \end{center}}
        \else
             {\begin{center}
             {\footnotesize Table~\thetable. #1}
              \end{center}}
        \fi}

\def\@citex[#1]#2{\if@filesw\immediate\write\@auxout
	{\string\citation{#2}}\fi
\def\@citea{}\@cite{\@for\@citeb:=#2\do
	{\@citea\def\@citea{,}\@ifundefined
	{b@\@citeb}{{\bf ?}\@warning
	{Citation `\@citeb' on page \thepage \space undefined}}
	{\csname b@\@citeb\endcsname}}}{#1}}

\newif\if@cghi
\def\cite{\@cghitrue\@ifnextchar [{\@tempswatrue
	\@citex}{\@tempswafalse\@citex[]}}
\def\citelow{\@cghifalse\@ifnextchar [{\@tempswatrue
	\@citex}{\@tempswafalse\@citex[]}}
\def\@cite#1#2{{$\null^{#1}$\if@tempswa\typeout
	{IJCGA warning: optional citation argument 
	ignored: `#2'} \fi}}

\def\pmb#1{\setbox0=\hbox{#1}
	\kern-.025em\copy0\kern-\wd0
	\kern.05em\copy0\kern-\wd0
	\kern-.025em\raise.0433em\box0}

\def\fnt#1#2{\footnotetext{\kern-.3em
	{$^{\mbox{\scriptsize #1}}$}{#2}}}

\def\fpage#1{\begingroup
\voffset=.3in
\thispagestyle{empty}\begin{table}[b]\centerline{\footnotesize #1}
	\end{table}\endgroup}

\def\runninghead#1#2{\pagestyle{myheadings}
\markboth{{\protect\footnotesize\it{\quad #1}}\hfill}
{\hfill{\protect\footnotesize\it{#2\quad}}}}
\font\tenrm=cmr10
\font\tenit=cmti10 
\font\tenbf=cmbx10
\font\bfit=cmbxti10 at 10pt
\font\ninerm=cmr9

\font\eightrm=cmr8

\def\qed{\hbox{${\vcenter{\vbox{			
   \hrule height 0.4pt\hbox{\vrule width 0.4pt height 6pt
   \kern5pt\vrule width 0.4pt}\hrule height 0.4pt}}}$}}

\begin{document}

\runninghead{Search for a Higgs boson decaying into two photons
$\ldots$} {Search for a Higgs boson decaying into two photons
$\ldots$}

\normalsize\textlineskip
\thispagestyle{empty}
\setcounter{page}{1}

\copyrightheading{}			

\vspace*{0.88truein}

\fpage{1}
\centerline{\bf SEARCH FOR A HIGGS BOSON DECAYING INTO TWO PHOTONS}
\vspace*{0.035truein}
\centerline{\bf WITH THE L3 DETECTOR AT LEP}
\vspace*{0.37truein}
\centerline{\footnotesize AURA ROSCA}
\vspace*{0.015truein}
\centerline{\footnotesize\it Humboldt University, Berlin, Invalidenstr. 110}
\baselineskip=10pt
\centerline{\footnotesize\it D-10115 Berlin, Germany}
\baselineskip=10pt
\centerline{\footnotesize\it e-mail: ars@physik.hu-berlin.de}
\baselineskip=10pt
\centerline{\small On behalf of the L3 collaboration}
\vspace*{10pt}

\vspace*{0.21truein}
\abstracts{
A search is performed for a Higgs boson, decaying into two photons,
using the L3 data collected at centre of mass energies between $\sqrt
s$ = 189 and 202 GeV,
corresponding to an integrated luminosity of 400 pb$^{-1}$. The processes
$e^{+}e^{-}\to \mbox{Zh} \to \mbox{q} \bar{\mbox{q}} \gamma \gamma $,
$e^{+}e^{-}\to \mbox{Zh} \to \nu \bar{\nu} \gamma \gamma $,
$e^{+}e^{-}\to \mbox{Zh} \to \LL \gamma \gamma $ are
considered.
The
observed
data are found to be consistent with the expected background from standard
physics processes. Limits on the branching fraction of the Higgs boson decay 
into two photons as a function of the Higgs mass are shown and a lower mass 
limit on a fermiophobic Higgs is derived.
}{}{}

\textlineskip			
\vspace*{12pt}			

\vspace*{1pt}\textlineskip	
\noindent
The Standard Model (SM) has been tested to a very good accuracy at LEP, SLC,
the Tevatron 
and at HERA. However, in spite of its great success, the
mechanism to generate the particle masses 
has not been verified experimentally. In the Standard Model the vector bosons
and fermion masses are generated through their interactions with the 
Higgs field which should manifest itself as a neutral spin-0 boson,
the Higgs boson. No direct observation of a Higgs boson has been made yet
and current direct searches constraint its mass to $M_{\mbox{h}}$ $>$ 113 GeV \cite{4}
in the minimal SM. 
Here we investigate models with extended Higgs sectors. In this
contribution we report on the
search for a Higgs boson produced in association with a Z boson
through the process $\mbox{e}^{+}\mbox{e}^{-}\to$Zh, followed by the decay
h$\to \gamma \gamma$, in all decay modes of the Z boson. In the Standard Model,
the decay of a Higgs boson h into a photon pair occurs via a
quark- or W-boson loop and its branching fraction is small.
However, several extended models predict enhancements of this branching fraction.
In Two Higgs Doublet Models of Type I \cite{1}, with an appropriate choice of
the model parameters, the lightest CP even Higgs boson does not couple to
fermions at tree level. Such a Higgs is expected to decay dominantly
into a pair of photons if its mass is below 90 GeV.

\noindent
The data were collected with the L3 detector at a centre-of-mass energy
between $\sqrt s$ = 189 and 202 GeV and amount to an integrated luminosity 
of about 400 pb$^{-1}$.

\noindent
A cut-based selection is performed in order to select events with photons
and to identify the Z in its various decay modes. This gives rise to
$\mbox{q}\bar{\mbox{q}}\gamma \gamma$, 
$\nu\bar{\nu}\gamma \gamma$
and $\ell^+ \ell^- \gamma \gamma$ final states, 
with $\ell$ =
e,$\mu$,$\tau$. The selection criteria for each final state rely
on a common photon identification and will be briefly summarised in the
following. The interested reader should consult Ref. \cite{2} for more 
details.

\noindent
Photons are identified as clusters in the electromagnetic calorimeter with a
shower shape compatible with that of an electromagnetic shower and which are
not associated with a charged track. To suppress photons from initial
state radiation, the most serious background for this search, only photons
in the polar angle ranges
$45^{\circ}< \theta <135^{\circ}$, $25^{\circ}< \theta <35^{\circ}$ 
or $145^{\circ}< \theta <155^{\circ}$ are accepted. To ensure that the
pair of photons arise from the decay of a heavy resonance we require the
energy of the most energetic photon to be larger than 10 GeV and the
energy of the second most energetic photon to be larger than 6 GeV.

\noindent
The signature for the $\mbox{q}\bar{\mbox{q}}\gamma \gamma$ final state
is a pair of isolated photons accompanied by two jets. To select these
events we apply a hadronic preselection requiring high multiplicity
events with large visible energy in the detector. Furthermore, only
those events which contain at least two photons are selected. 
The recoil mass against the di-photon system must be consistent with the Z mass,
$|M_{\rm recoil}-M_{\rm Z}|<15$ GeV. 

\begin{table}[t]
\tcaption{Number of events expected from Standard Model processes
compared to the observed number of events, together with the signal efficiency
for a hypothetical Higgs boson of  95 GeV mass. }
\centerline{\footnotesize\smalllineskip
\begin{tabular}{ccccccc}
\hline
&\multicolumn{3}{c}{$\sqrt{s}=189$ GeV, $L_{int}=176$ pb$^{-1}$} &
\multicolumn{3}{c}{$\sqrt{s}=192-202$ GeV, $L_{int}=233$ pb$^{-1}$}\\
Final State &Data&Bkgd. predicted&Efficiency&Data&Bkgd.
predicted&Efficiency\\ \hline
$\mbox{q}\bar{\mbox{q}}\gamma \gamma$  & 10 & 16.2 & 43$\%$ & 14 & 12.6 & 45$\%$ \\
 $\nu{\bar{\nu}}\gamma \gamma$    &\phantom{0} 3&\phantom{0} 4.3 & 29$\%$ & - & - & -  \\
$\ell^+ \ell^- \gamma \gamma$    &\phantom{0} 5&\phantom{0} 2.5 & 37$\%$ & - & - & - \\ \hline
\end{tabular}}
\end{table}

\noindent
The  $\nu{\bar{\nu}}\gamma \gamma$ final state is characterised
by the presence of two photons and missing energy in the event. To reduce 
the background from the $\mbox{e}^{+}\mbox{e}^{-}\to\gamma \gamma (\gamma)$
process and from double radiative events with final state particles escaping
detection, we require photon acoplanarity and a total transverse momentum
of the di-photon system greater than 3 GeV. The missing mass must be consistent
with the Z boson mass within $\pm$10 GeV. 

\noindent
The $\ell^+ \ell^- \gamma \gamma$ final state is
characterised by the presence of two photons and a pair of same type leptons
in the event. We require the recoil mass against the di-photon system 
to be consistent with the Z mass,
$|M_{\rm recoil}-M_{\rm Z}|<15$ GeV. 

\noindent
The number of data and expected
background events left after the selection is applied, as well as the signal
efficiency in each of the studied channels, are reported in Table 1.  

\noindent
Since no signal is observed in the data, we evaluate the confidence level 
for the absence of a Higgs signal using the reconstructed di-photon invariant
mass as a final discriminant variable. This distribution is shown in Figure 1(a)
for all Z final states combined.

\begin{figure}[t]
\centerline{
\begin{minipage}[t]{6.0cm}
\vspace*{13pt}
\centerline{\hspace{-0.6cm}
\epsfig{file=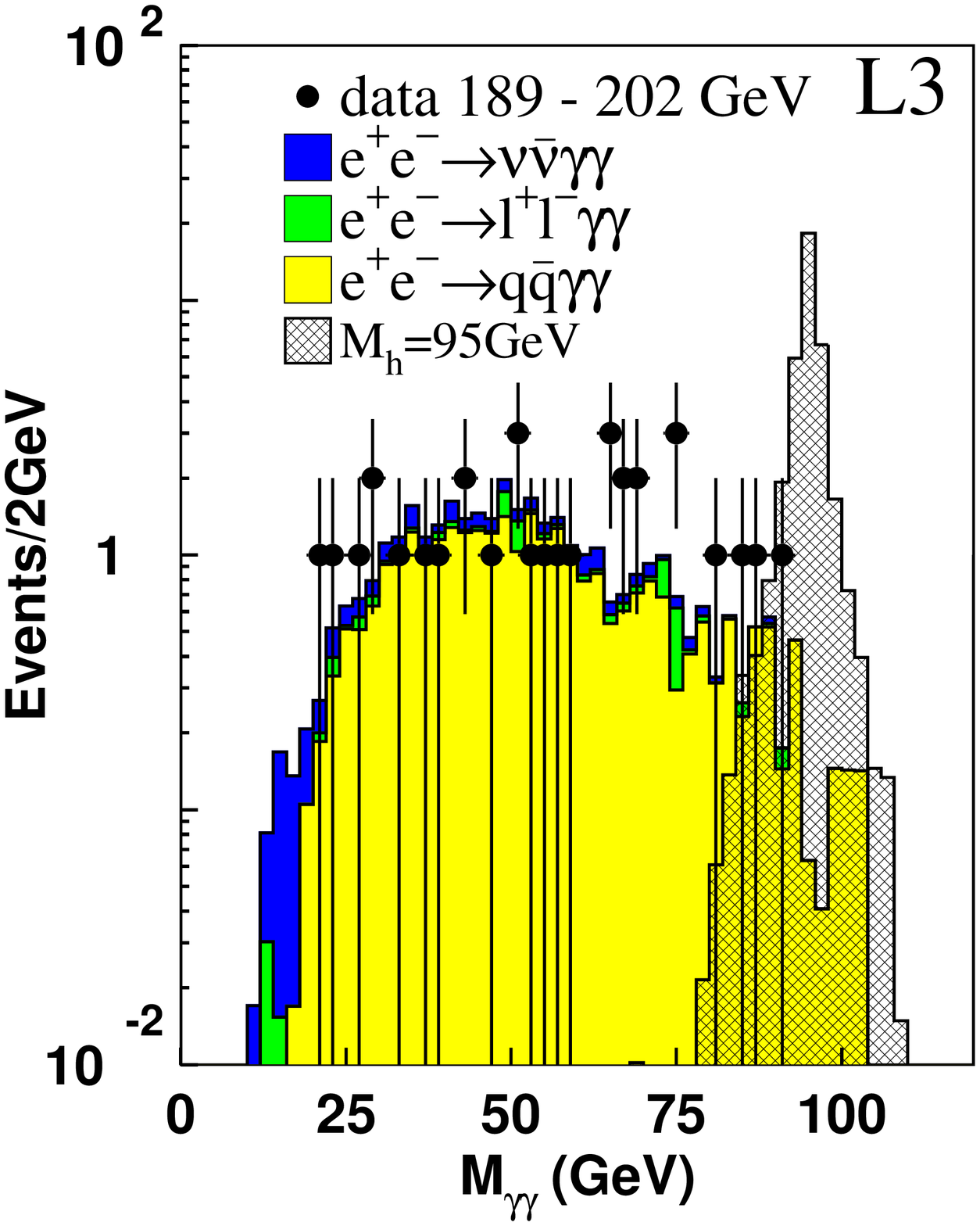,width=6.1cm,clip=}}
\centerline{(a)}
\vspace*{13pt}
\end{minipage}
\hspace*{0.5cm}
\begin{minipage}[t]{6.0cm}
\vspace*{13pt}
\centerline{\hspace{-0.2cm}
\epsfig{file=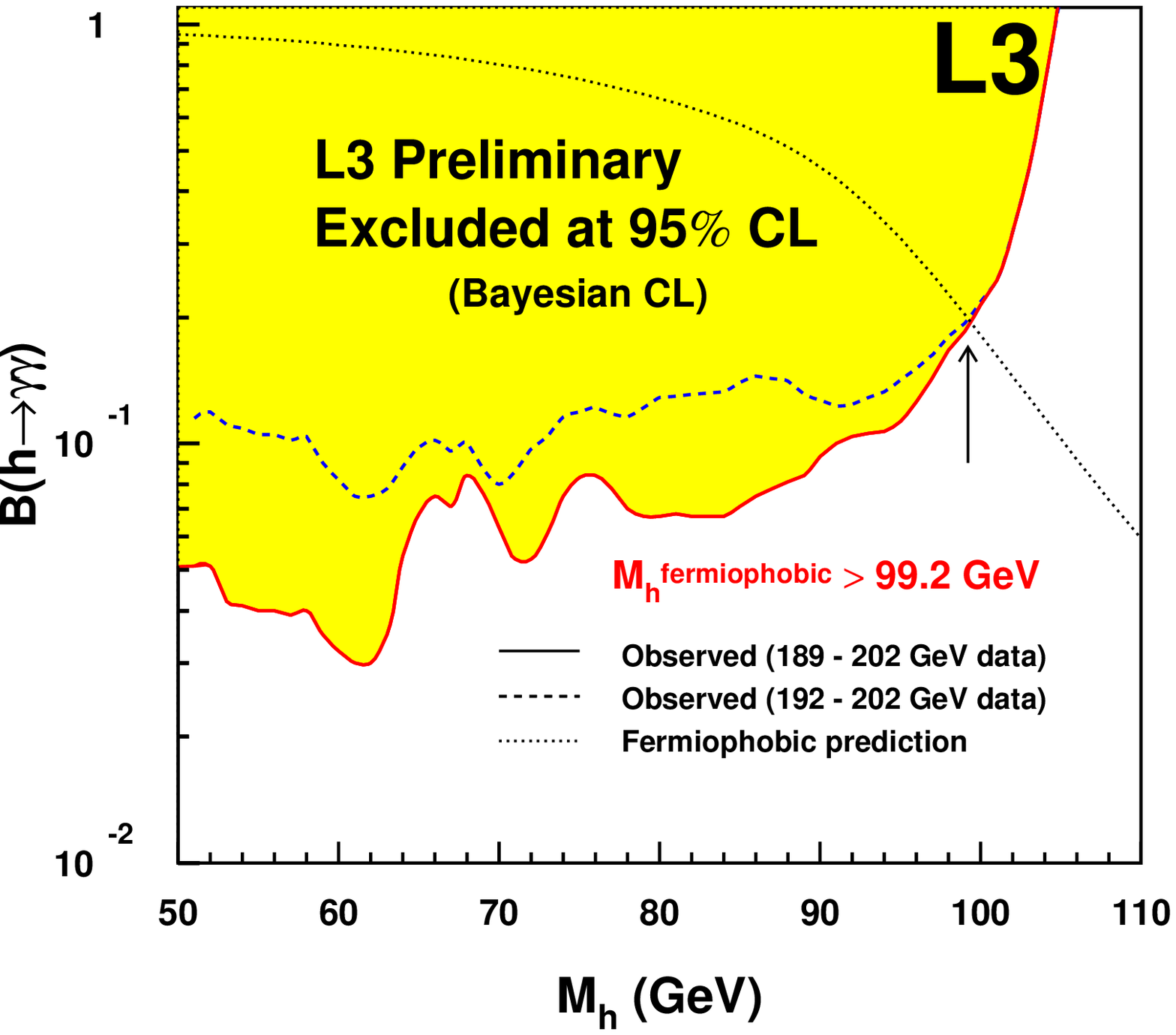,width=7.3cm,clip=}}
\centerline{(b)}
\vspace*{13pt}
\end{minipage}
}
\fcaption{
(a) The distribution of the reconstructed di-photon invariant mass
for all Z final states combined, after the final selection, in data,
background and for a Higgs boson signal with the mass 
$M_{\mbox{h}}$ = 95 GeV. The signal, assuming the Standard Model
cross section and a branching fraction B(h $\to \gamma \gamma$) = 1, is
superimposed and normalised to the integrated luminosity.  
(b) Excluded values of the branching fraction B(h $\to \gamma \gamma$) as a function
of the Higgs mass, in the assumption of a Standard Model production
cross section. The theoretical prediction is also presented. 
}
\end{figure}

\noindent
The calculation of the limits takes into account
systematic uncertainties of 1$\%$ from the signal Monte Carlo statistics,
1.5$\%$ from the simulation of the photon isolation criteria and 4$\%$ on the
number of expected background events. Figure 1(b) shows the measured upper limits
on the branching fraction
B(h$\to \gamma \gamma$) as a function of the Higgs mass assuming the
Standard Model rate for the Zh production. The theoretical prediction is also
shown for a fermiophobic Higgs boson as calculated with the HDECAY program \cite{3}. The
lower limit on the mass of a fermiophobic Higgs boson is set at  
$M_{\mbox{{\scriptsize h}}} > 99.2$ GeV at 95$\%$ confidence level.
The LEP Higgs Working Group has combined the data up to $\sqrt s$ = 209 GeV from 
the four
experiments and obtained a lower mass limit of 106.4 GeV at 95$\%$ confidence level \cite{4}.

\nonumsection{Acknowledgements}
\noindent
I thank the German Bundesministerium f\"ur Bildung und Forschung for its 
financial support.
 
\nonumsection{References}

\end{document}